\documentclass[epj]{svjour}
\usepackage{graphicx}
\usepackage{epstopdf}
\usepackage{dcolumn}
\usepackage{bm}
\usepackage{epsfig}
\usepackage{amsfonts}
\usepackage{amssymb,amscd}
\usepackage{hyperref}
\usepackage{xcolor}
\hypersetup{
    colorlinks=true,                
    breaklinks=true,                
    urlcolor= black,                
    linkcolor= blue,                
    bookmarksopen=false,
    filecolor=black,
    citecolor=blue,
    linkbordercolor=blue,
    pdfborder = {0 1 0}
}

\begin{document}

\title{A study on the isolated photon production in nuclear collisions at the CERN-LHC energies}

\author{
G. Sampaio dos Santos\inst{1} \and
G. Gil da Silveira\inst{1,2} \and
M. V. T. Machado\inst{1}
}

\institute{Grupo de Fenomenologia de Partículas de Altas Energias, GFPAE  IF-UFRGS \\
Caixa Postal 15051, CEP 91501-970, Porto Alegre, RS, Brazil \and
Departamento de F\'{\i}sica Nuclear e de Altas Energias, Universidade do Estado do Rio de Janeiro\\
CEP 20550-013, Rio de Janeiro, RJ, Brazil}

\date{Received: date / Revised version: date}

\abstract{
An analysis of prompt photon production in high energy nuclear collisions at the LHC energy regime is performed within the parton saturation picture taking into account the updated phenomenological color dipole models. The results are confronted with the measurements made by the ALICE, ATLAS, and CMS experiments in terms of photon transverse momentum at different rapidity bins. As a result, we show that the prompt photon production exhibits distinct scalings in $AA$ events associated to geometrical properties of the collision and can be properly addressed in the color dipole formalism.
}


\authorrunning{Sampaio dos Santos, Gil da Silveira, and Machado}
\titlerunning{A study on the isolated photon production in nuclear collisions...}
\maketitle

\maketitle

\section{Introduction}
\label{intro}

The hard electromagnetic probes, such a photon, are a powerful tool to investigate the hadronic and nuclear interactions, especially in the high-energy regime due to their scanning features. As they interact only electromagnetically and are colorless objects, their passage through dense hadronic matter are not disturbed, like in a quark-gluon deconfined medium known as Quark-Gluon Plasma (QGP). Hence, photons are produced at any time of the collision and cross the interaction path unaffected, providing information on the scenario lead to their production. Photons produced directly from the hard scattering of partons are considered prompt photons, differently from those originated by the decay of a hadron, such as $\pi^{0}$ or $\eta$. However, photons originate from distinct sources and production mechanisms, hence an isolation requirement is applied to reduce the effect of sizable backgrounds. Experimentally, isolation cuts establish an energy threshold in the vicinity of the photon, identified as a cone of radius $R$ in rapidity $y^{\gamma}$ and azimuthal angle $\phi^{\gamma}$ around the photon direction. Here we pay special attention to the production of hard isolated photons, providing a clean probe to investigate the QCD dynamics \cite{pasechnik,acharya,gordon1,frixione}. The elementary diagrams associated with the underlying processes are theoretically well known and contributions from fragmentation processes can be suppressed due to the imposition of an isolation criteria \cite{ichou}.

The WA98 Collaboration \cite{aggarwal} was the first experiment to report results on direct photon production in heavy-ion collisions with the measurement of the prompt photon yield in central PbPb collisions at 17.3~GeV within the transverse momentum range of \mbox{0.5~$<p_T<$~4~GeV/c}. At the LHC it has been possible to investigate prompt photons in $AA$ collisions covering a larger center-of-mass energy plus wide windows of rapidity and transverse photon momentum distributions.

Typically, the evaluation of $AA$ collisions involves the determination of the geometrical overlap area of the nuclear targets at a certain range of the collision impact parameter, which is connected to the centrality class. Centrality is defined in terms of the fraction of the total inelastic hadronic cross section that has been measured and corresponds to percentile values, where 0\% and 100\% indicates the most central and peripheral collisions, respectively. These intervals are related to the geometrical quantities present in the collisions and are useful for the analysis of the heavy-ion events. Often such geometric properties are determined by Monte Carlo implementation of the Glauber model \cite{miller} taking into account the impact parameter and the nucleon density distribution in the nucleus.

From the experimental side it was verified different forms of scalings in the $p_T$ spectrum of direct photon production in heavy-ion and hadronic reactions, namely multiplicity and geometrical scalings. The first is related to the charged hadron pseudo-rapidity density at midrapidity \cite{kbml,pras,kp} while the latter implies that the cross sections for photon-target processes are a function of a dimensionless single scaling variable \cite{Stasto:2000er,munier}. Here we will study prompt photon production in $AA$ collisions using two scalings associated to the geometric properties of the reactions.

The prompt photon process can be studied in the target rest system, which resembles an electromagnetic quark (antiquark) bremsstrahlung \cite{kop,kop1,kop2} and, accordingly to the QCD dipole picture, is described in terms of $q\bar{q}$ dipole scattering off the target. In such framework the phenomenology is based on the dipole cross section that accounts the nonlinear gluon recombination effect and adjusted to the DIS data to successfully describe the measurements of inclusive and exclusive processes. The gluon density dynamics is interconnected with a transition region limited by a $x$-dependent saturation scale, $Q_s(x)$, namely the transverse momentum scale at which the gluon density is tamed as expected from the nonlinear behavior of the QCD evolution. Moreover, this dense and saturated region at low Bjorken-$x$ is expected to be accessed with  measurements of high-energy prompt photon production.

In this work predictions are performed for prompt photon yields focusing in heavy-ion collisions at the LHC energies taking into account large and low $p_T$-spectra at different rapidity bins. For other studies concerning prompt photons within the color dipole picture and others approaches see, e.g., Refs.~\cite{Benic:2018hvb,Ducloue:2017kkq,Rezaeian:2009it,Krelina:2016hkr,JalilianMarian:2012bd,Kovner:2014qea,Rezaeian:2016szi,Benic:2017znu,vic,goharipour,campbell,roy,gsds,gsds1}. Clearly understanding the hard probes as direct photon in heavy-ion collisions is crucial to single out the underlying properties of the quark-gluon plasma. Particularly, we will investigate two different scaling proposals: the $N_{part}$ scaling based on parton saturation arguments and the usual $N_{coll}$ scaling from Glauber model applied to QCD hard processes.

The paper is organized as follows. In Sec.~\ref{dirph} we start by providing the theoretical framework presenting the main expressions to obtain the prompt photon yields in the color dipole framework.  In Sec.~\ref{res} we show our predictions evaluated in view of the experimental measurements at the LHC reported by the ALICE, ATLAS, and CMS Collaborations. Finally, we presented the main conclusions and remarks in Sec.~\ref{conc}.

\section{Theoretical formalism}
\label{dirph}

At high energies the color dipole formalism is applied to investigate radiation process, for instance, the real photon production off protons and nuclear targets. In such scenario, a prompt photon is emitted via an electromagnetic bremsstrahlung by a quark (antiquark) projectile that exchange a single gluon with the color field of the target. Therefore, the real photon production process can be configured as a color dipole scattering off the proton/nucleus. In Ref.~\cite{kop3} the differential cross section for prompt photon production is derived taking the proton as target, being obtained by convoluting the proton structure function with the partonic cross section,
\begin{eqnarray}\nonumber
\frac{d^3\sigma\,(pp\to \gamma
X)}{dy^{\gamma}d^{2}\vec{p}_{T}} &=&
\frac{\alpha_{em}}{2\pi^2}\int_{x_{1}}^{1}\frac{d\alpha}{\alpha}
 F_{2}^{(P)}\left(\frac{x_{1}}{\alpha},\mu^2\right) \left\{ m_q^2\alpha^4 \right.\\
 &\times& \left.\left[\frac{{\cal I}_1}{(p_T^2+\varepsilon^2)}-\frac{ {\cal I}_2}{4\varepsilon} \right]
  +  [1+(1-\alpha)^2]\right.\nonumber \\
  &\times& \left. \left[ \frac{\varepsilon p_T \, {\cal I}_3}{(p_T^2+\varepsilon^2)} -\frac{{\cal I}_1}{2}+\frac{\varepsilon \,{\cal I}_2}{4}\right]
\right \},
\label{hank}
\end{eqnarray}
where $y^{\gamma}$ and $p_{T}$ are the photon rapidity and transverse momentum, respectively. Furthermore, $F_{2}^{(P)}$ denotes the structure function of the projectile ($P$) particle and the Hankel integral transforms of order ${\cal{O}}(1)$ corresponding to ${\cal I}_{1,2}({\cal I}_{3}$) are written as
\begin{eqnarray}
{\cal I}_1 & = & \int_0^{\infty}dr\,rJ_0(p_T\,r)K_0(\varepsilon\,r)\,\sigma_{dip}(x_2,\alpha r), \label{hankel1} \\
{\cal I}_2  &=&  \int_0^{\infty}dr\,r^2J_0(p_T\,r)K_1(\varepsilon\,r)\, \sigma_{dip}(x_2,\alpha r), \label{hankel2} \\
{\cal I}_3 & = & \int_0^{\infty}dr\,rJ_1(p_T\,r)K_1(\varepsilon\,r)\, \sigma_{dip}(x_2,\alpha r).
\label{hankel3}
\end{eqnarray}
In the expressions above $\alpha$ represents the relative fraction of the quark momentum exchanged in the quark-photon vertex, while $x_{1,2}$ are the Bjorken variables given by $x_{1,2} = \frac{p_T}{\sqrt{s}}e^{\pm y^{\gamma}}$, with $\sqrt{s}$ being the collision center-of-mass energy. Moreover, the Hankel transforms have the auxiliary variable $\epsilon^{2}=\alpha^{2} m_{q}^{2}$ which depends on the effective quark mass, which are assumed to be $m_{q}=0.2$~GeV in our calculations.  

As discussed before the prompt photon production mechanism can be seen as an effective $q\bar{q}$ dipole interacting with the target, where the dipole cross section, $\sigma_{dip}$, contains all the information about the strong interaction dynamics. In this work we will consider phenomenological models which account for the gluon saturation and present the following features: (i) for large dipole transverse sizes, $r$, $\sigma_{dip}$ saturates, namely $\sigma_{dip} \rightarrow \sigma_0$; (ii) for the opposite case, small dipole sizes, one has 
$\sigma_{dip} \sim r^2$, that is, the dipole cross section vanishes as expected in the color transparency phenomenon \cite{kop4}. Typically, the dipole-proton cross section assumes the parameterized form
\begin{eqnarray}
\label{sigdip}
\sigma_{dip}(x,\vec{r};\gamma) &=&\sigma_{0}\left[ 1-\exp\left(-\frac{r^{2}Q_{s}^{2}}{4}\right)^{\gamma_{\mathrm{eff}}}\,\right], \label{dip} \\ Q_{s}^2(x) &=& \left(\frac{x_0}{x}\right)^{\lambda},
\label{param}
\end{eqnarray}
where the effective anomalous dimension is denoted by $\gamma_{\mathrm{eff}}$ whereas $Q_{s}$ stands for the saturation scale. We will employ the GBW model \cite{gbw}, which takes $\gamma_{\mathrm{eff}} = 1$ and the recent fitting parameters from DIS data at DESY-HERA collider $\sigma_0 = 27.32$~mb, $x_0 = 0.42\times 10^{-4}$, and $\lambda = 0.248$ \cite{gbwfit}. The BUW model \cite{buw} also considers the cross section established in Eq.~(\ref{dip}), however the effective anomalous dimension reads
\begin{eqnarray}
\gamma_{\mathrm{eff}}= \gamma_{s}+(1-\gamma_{s})\frac{(\omega^a-1)}{(\omega^a-1)+b},
\label{buw}
\end{eqnarray}
with $\omega\equiv p_T/Q_{s}$ and the free parameters are fitted in order to describe the RHIC data on hadron production $a=2.82$ and $b=168$. The remaining parameters are $\gamma_{s} = 0.63$, $\sigma_0 = 21$~mb, $x_0 = 3.04\times 10^{-4}$, and $\lambda = 0.288$. Additionally, we will include into the analyses the Impact Parameter Saturation (IPSAT) model \cite{ipsat,ipsatfit} that includes the QCD evolution effects in the dipole cross section,
\begin{eqnarray} 
\sigma_{dip} (x,\vec{r}) &=& 2\int d^2b\,N(x,r,b),\\
N(x,r,b)& = & 1-\exp\left(-\frac{\pi^2}{2N_c}r^2\alpha_S(\mu^2)xg(x,\mu^2)T(b)\right).\nonumber
\label{ipsat1}
\end{eqnarray}
Now, the dipole cross section accounts for a gluon distribution evolved via DGLAP evolution equations and an impact-parameter dependence is encoded by a Gaussian profile for the proton,
\begin{eqnarray}
T(b) = \frac{1}{2\pi B_{G}}  
\exp\left(-\frac{b^{2}}{2B_{G}} \right).
\label{profile}
\end{eqnarray}
We can obtain analytically the Hankel integrals in the color transparency region, Eqs.~(\ref{hankel1}--\ref{hankel3}),
\begin{eqnarray}
{\cal I}_1  &\propto& \frac{(\varepsilon^2-p_T^2)}{(p_T^2+\varepsilon^2)^3},\\
{\cal I}_2 &\propto& \frac{4\varepsilon\,(\varepsilon^2-2p_T^2)}{(p_T^2+\varepsilon^2)^4},\\
{\cal I}_3 &\propto& \frac{2p_T\varepsilon}{(p_T^2+\varepsilon^2)^3},
\label{aproxis}
\end{eqnarray}
following the Ref.~\cite{mm} where the exact prefactors can be identified using the GBW model. We will use this particular case latter.

Commonly, in nucleus-nucleus collisions the observables are determined in a given centrality class. Hence, geometrical quantities associated to the collision centrality are useful in order to analyze the observables, such as: number of binary nucleon-nucleon collisions, $N_{coll}$, number of participant nucleons, $N_{part}$, and geometric nuclear overlap function, $T_{AA}$. Generally, such quantities are defined from the Glauber model and calculated via Monte Carlo methods. Besides, for the purpose of evaluating the prompt photon production in $AA$ collisions, it is necessary to take into consideration the nuclear structure function $F_{2}^{A}$ that enters in Eq.~(\ref{hank}). One source of the theoretical uncertainties in this formalism is related to the scale $\mu^{2}$, identified as $\mu^{2} = p^{2}_{T}$, in our numerical calculations. Following Ref.~\cite{f2a} we use a parametrization for $F_{2}^{A}$ (fortran code) presented there, which is obtained by applying the leading twist model of nuclear shadowing with DGLAP evolution for the nuclear PDFs.

In nuclear high-energy collisions the QCD nuclear effects take place, in particular those originated from multiple parton scattering and nonlinear gluon recombination. The nuclear effects can be essentially evaluated within the color dipole framework in two forms: geometric scaling property derived from parton saturation models and a Glauber-Gribov approach for nuclear shadowing. In this work, we will assume the geometric scaling in the dipole-nucleus amplitude, $N_A$. The geometric scaling reflects the fact that the nuclear effects are absorbed into the saturation scale and on the transverse area of the colliding nuclei. In Ref.~\cite{salgado} the proposed scaling establishes the dependence on $A$ in the scattering cross section and the nuclear effects are embedded onto the nucleus transverse area in relation to the proton one ($S_A=\pi R_A^2$ and $S_p=\sigma_0/2=\pi R_p^2$, where $R_A \simeq 1.12 A^{1/3}$~fm is the nucleus radius).
The saturation scale acquires a dependence on the collision energy, transverse momentum and collision centrality class \cite{kbml,pras,kp} and the latter is characterized by the number of participants. Then, $Q_{s,p}$, the proton saturation scale, is replaced properly by a $N_{part}$-dependent $Q_{s,A}$, the correspondent nuclear saturation scale version (see Ref.~\cite{kp} and references therein), 
\begin{eqnarray}
Q_{s,A}^2&=&Q_{s,p}^2\left(\frac{\kappa(b)\,N_{part}\, \pi R_p^2}{\pi R_A^2}\right)^\frac{1}{\delta},  \label{qs2A} \\
N_A(x,r,b) & = & N(rQ_{s,p}\rightarrow rQ_{s,A}),
\label{qs2AN}
\end{eqnarray}
where $\kappa(b)$ stands for a parameter dependent on the impact parameter $b$ and the values $\delta = 0.79$ and $\pi R_p^2=1.55$~fm$^2$ have been fitted to data \cite{salgado}. In particular, we are assuming the scaling on the number of participants proposed in \cite{kbml,pras,kp} and the parametrization for the nuclear saturation scale given in Ref.~\cite{salgado}.
Therefore, the assumptions above are translated into the cross section for prompt photon production in $AA$ collisions accordingly to
\begin{eqnarray}
\frac{d^3\sigma(AA\to \gamma X)}{dyd^2\vec{p}_T} = \left(\frac{S_A}{S_p}\right)\left.\frac{d^3\sigma(Ap\to \gamma X)}{dyd^2\vec{p}_T}\right|_{ Q_{s,p}^2\rightarrow Q_{s,A}^2},
\label{prescr}
\end{eqnarray}
where additional nuclear effects occur also in the nuclear structure function, $F_2^A \sim AF_2^p$, appearing in Eq.~(\ref{hank}). It should be stressed that the prescription above is an oversimplification and it should be taken with a grain of salt.

Alternatively, the photon spectrum in $AA$ collisions at a certain centrality class ($C_1 - C_2$) is possible to obtain by applying the correspondent $N_{coll}$ scaling. As a result, the $AA$ yield for this hard process using Glauber model is written  as \cite{dEnterria},
\begin{eqnarray}\nonumber
\frac{d^3\mathrm{N}\,(AA\to \gamma X)_{(C_1 - C_2)}}{dy^{\gamma}d^{2}\vec{p}_{T}}  = \big<N_{coll}\big>_{(C_1 - C_2)} \, \frac{d^3\mathrm{N}\,(pp\to \gamma X)}{dy^{\gamma}d^{2}\vec{p}_{T}}, \\
\label{Ncoll}
\end{eqnarray}
properly scaled for comparison to photons measured in $pp$ collisions (in our calculations we take the recent parametrization 
for $F_{2}^{p}$ in Ref.~\cite{block}) at the same energy. It is important to stress that there is an equivalence between the invariant yield ($\mathrm{N}$) and the cross section ($\sigma$) for some processes determined by $N_{AA} = \sigma_{AA}/\sigma^{geo}_{AA}$, where  
$\sigma^{geo}_{AA}$ is the geometrical $AA$ cross section.

In what follows we will employ the phenomenological models previously discussed to compute the direct photon yield in $AA$ collisions and performed a comparison with the corresponding experimental data from the LHC kinematic regime.

\section{Results and discussions}
\label{res}

\begin{figure*}[!t]
\begin{tabular}{cc}
\includegraphics[width=0.5\textwidth]{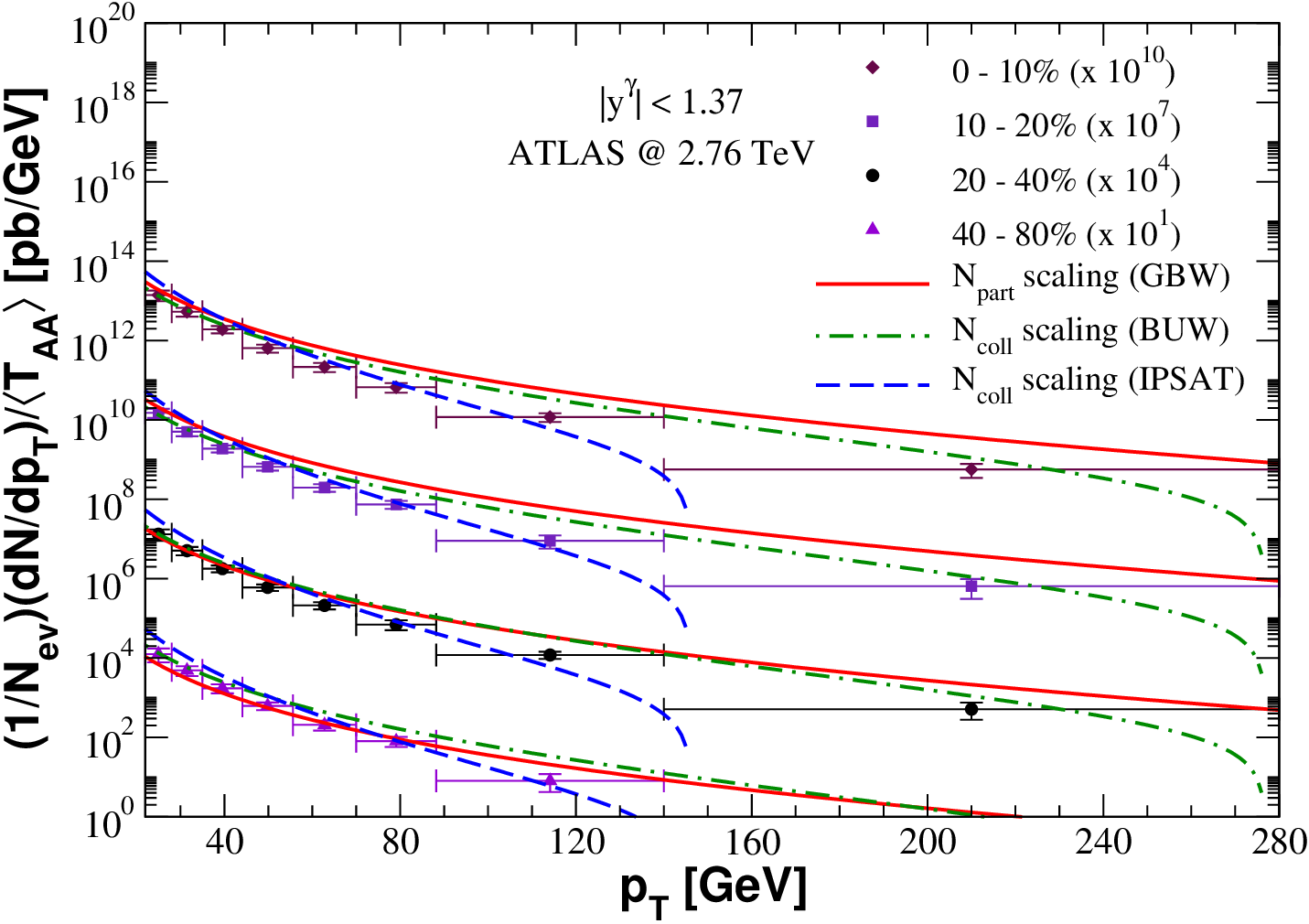}
\includegraphics[width=0.5\textwidth]{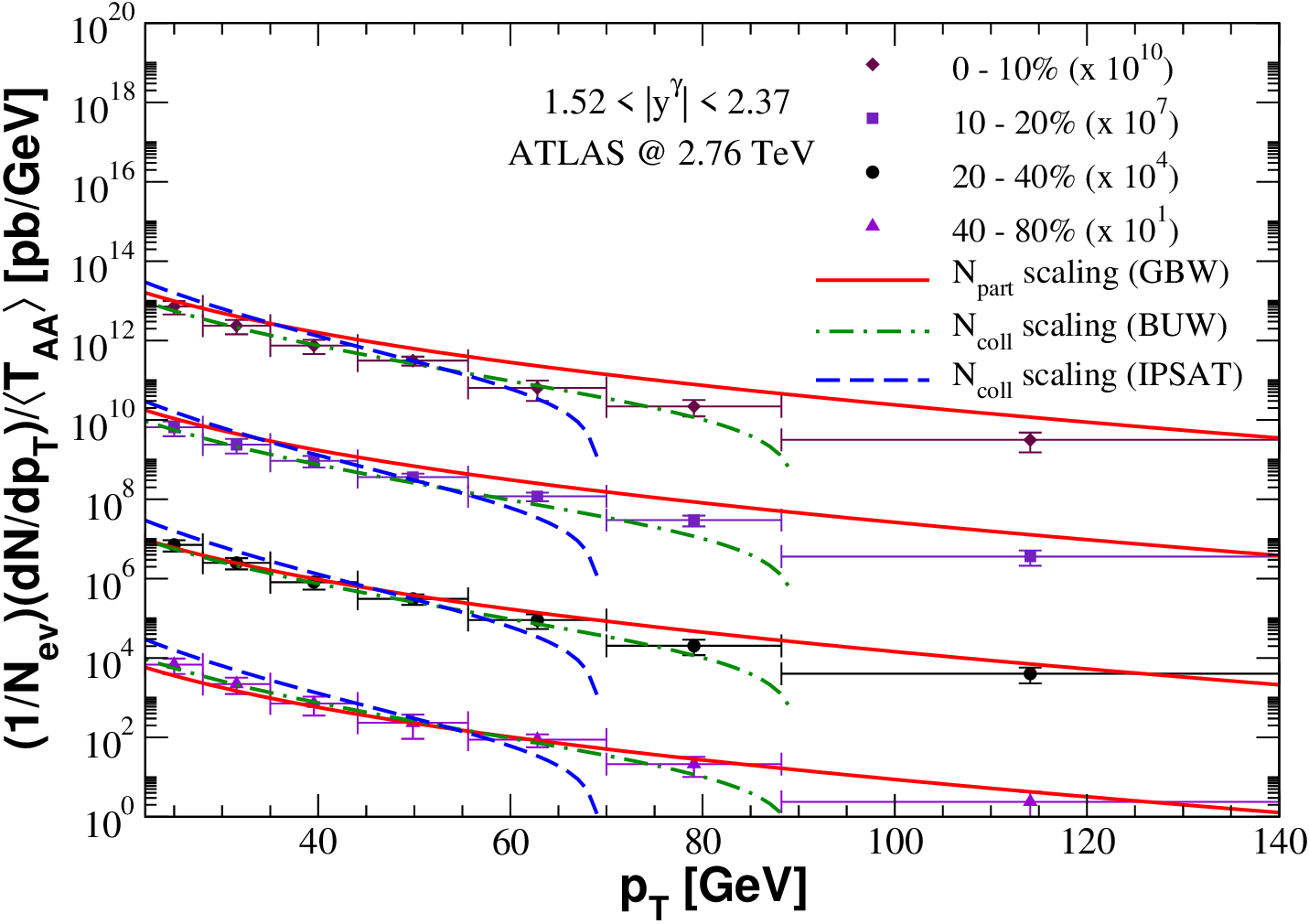}
\end{tabular}
\caption{The $p_T$ distribution for the prompt photon yield in PbPb collisions at $\sqrt{s} = 2.76$~TeV assuming particular photon rapidity bins. The results with two distinct scalings are compared to the experimental measurements reported by the ATLAS Collaboration \cite{atlas}.}
\label{atlas}
\end{figure*}

Here some comments are in order. We estimate the dipole-nucleus amplitude taking the $N_{part}$ and $N_{coll}$ scalings rules as input to our numerical calculations with the GBW and BUW/IPSAT models, respectively. Besides, we assume the small-$r$ limit for $\sigma_{dip}$ using IPSAT approach, which is convenient at the large $p_T$ region where $r \approx 1/p_T$ is sufficiently small in direct photon production and also the Hankel transforms can be solve analytically. Moreover, the color dipole formalism is particularly appropriate for small $x$ (or equivalent high-energy process), although it has a limit of validity of $x_2\leq10^{-2}$. However, Ref.~\cite{mm} has investigated the role played by large-$x$ corrections in prompt photon cross sections and it has to be considered in order to consistently describe the large-$x$ behavior of the associated phenomenology. Consequently, we have included this correction factor to our predictions by multiplying the GBW and BUW dipole cross sections by $(1-x_2)^n$ (with $n=7$). Concerning the IPSAT model, the threshold factor is already present at the initial scale for the gluon PDF parametrization. As a last consideration, an isolation cut is required as a experimental criteria. The color dipole model considers only the direct contribution concerning the prompt photon production and disregard the fragmentation contribution, which accounts to a contribution of 10\% at the LHC energies in midrapidities \cite{dEnterria1}. Therefore, very small modifications are expected in the $p_T$-spectrum due the isolation cut and this is implicitly assumed in the studies performed in Refs.~\cite{kop,kop5,kop6,Rezaeian:2009it,Krelina:2016hkr,vic,mm,kbml}.

Before showing the results, we inform that the values used for $N_{part}$, $N_{coll}$, and $T_{AA}$ are given in Refs.~\cite{pras,sirunyan,atlas,cms276,cms502,alice}. In Fig.~\ref{atlas} we present the predictions for the photon yield $p_T$ distribution in PbPb collisions compared to the data collected by the ATLAS experiment at $\sqrt{s} = 2.76$~TeV \cite{atlas} in two rapidity intervals. Considering all centrality class and the two rapidity bins at large $p_T$, we verified that the results with $N_{part}$ and $N_{coll}$ scalings indicate sizable differences, mostly at very large $p_T$. At lower transverse momentum both approaches produce similar results.

In the left panel of Fig.~\ref{atlas} are shown the results for central rapidities for different centralities, including the minimum bias one. The solid line stands for the $N_{part}$ scaling prediction, where the parameter $\kappa = \kappa(b)$ \cite{stasto} that accounts for the profile function of the nucleus in the nuclear saturation scale was set to $\kappa = 0.61$ and the nuclear structure function including leading twist shadowing was taken from Ref.~\cite{f2a}. For the dipole-proton amplitude, the phenomenological GBW model was considered. The data description is reasonable for all $p_T$. The dot-dashed and dashed lines represent the predictions from the $N_{coll}$ scaling considering two models for the dipole-proton cross section: BUW (dot-dashed curve) and IPSAT (dashed curve). The prediction from BUW is quite satisfactory whereas IPSAT shows important suppression at large $p_T$. In the right panel of Fig.~\ref{atlas} the experimental points correspond to forward rapidities. Concerning the models, they follow similar trend as in central rapidities, however the quality of data description for $N_{coll}$ scaling is degraded at large $p_T$. The $N_{part}$-scaling remains doing a good job in all $p_T$ range.
In order to illustrate the typical value of $x_2$, for instance by taking the more forward bin, $1.52 < |y^{\gamma}| < 2.37$, the smaller value probed is $x_2 \sim 8\times 10^{-4}$. Even at the central rapidities and large $p_T$, $x_2$ remains small being of order $\sim 0.07$. Finally, for $p_T < 40$~GeV the results for both scalings are similar and apparently it is not possible to discriminate between GBW e BUW predictions as they both are in  agreement with experimental measurements. One exception occurs for the most peripheral collisions (20--40\%, 40--80\%) where deviations between the results are visible.
 
\begin{figure*}[t]
\begin{tabular}{cc}
\includegraphics[width=0.5\textwidth]{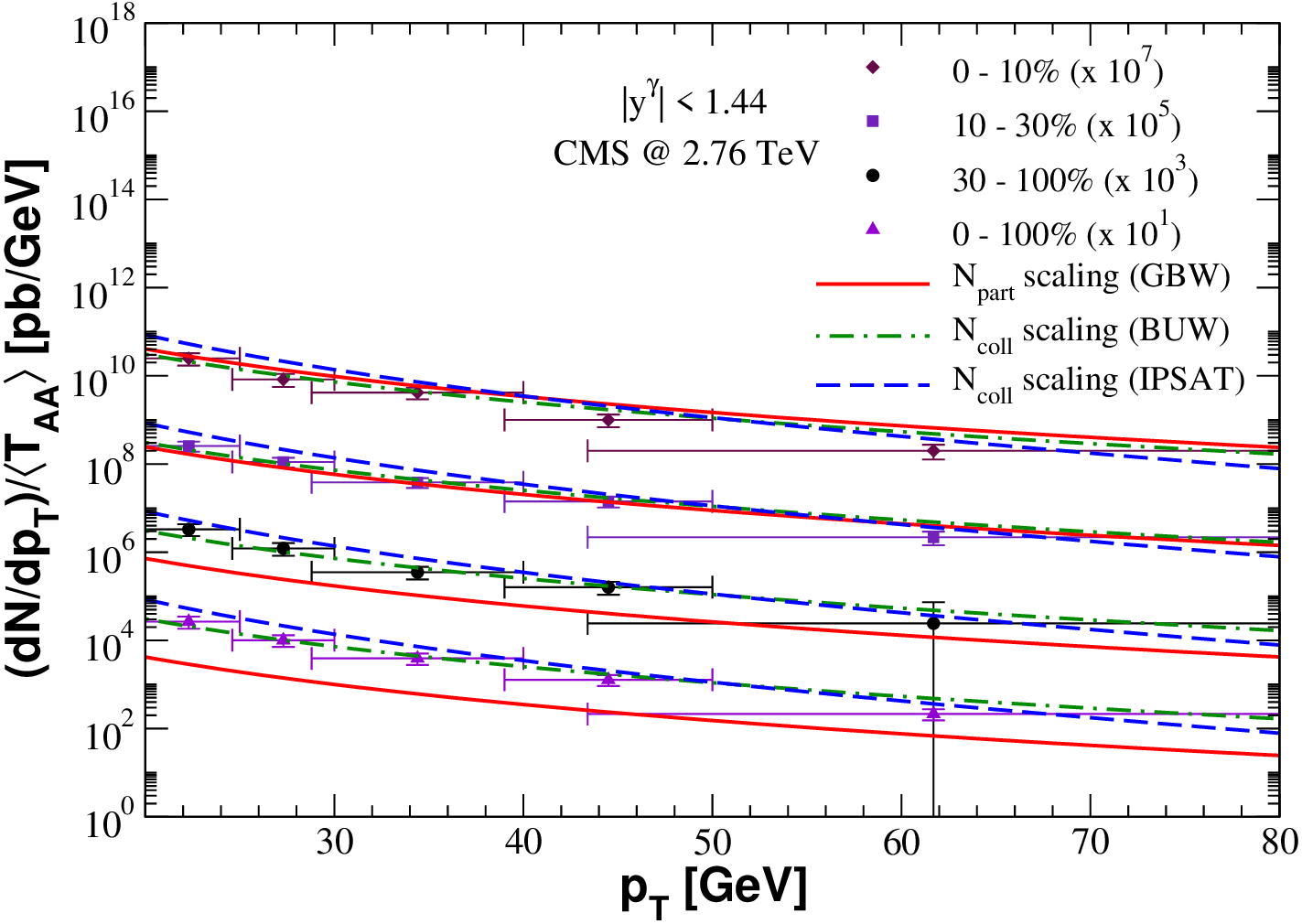}
\includegraphics[width=0.5\textwidth]{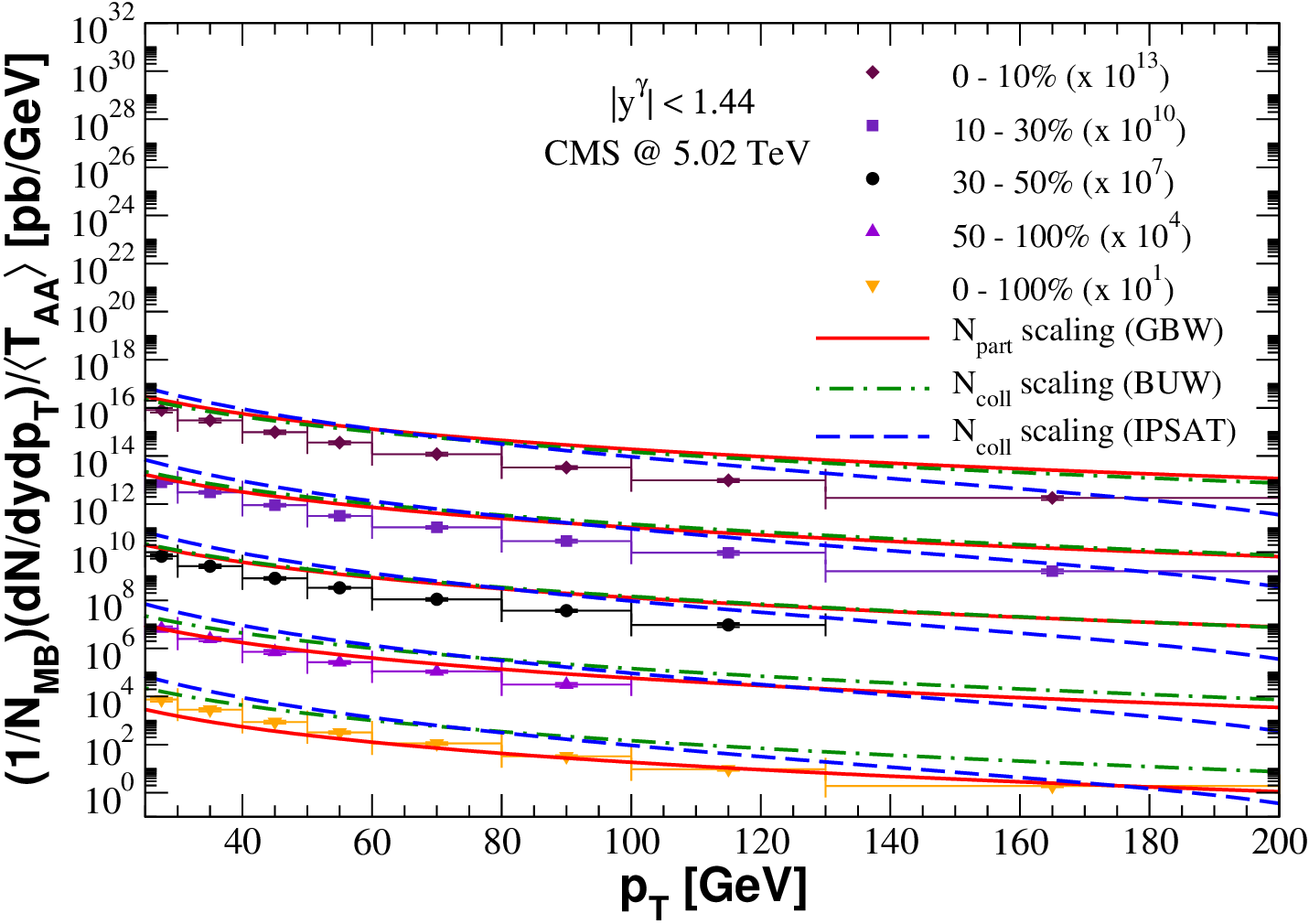}
\end{tabular}
\caption{The $p_T$ distribution for the prompt photon yield in PbPb collisions at $\sqrt{s} = 2.76$ and $5.02$~TeV assuming the same  photon rapidity bin. The results assuming two distinct geometrical scalings are compared to the experimental measurements reported by the CMS Collaboration \cite{cms276,cms502}.}
\label{cms}
\end{figure*}

In Fig.~\ref{cms} the results are compared to the experimental measurements reported by the CMS Collaboration \cite{cms276,cms502} considering the energies of $2.76$ (left panel) and $5.02$~TeV (right panel) at $|y^{\gamma}| < 1.44$. The notation for the curves is the same as the previous one. Here, one has a wide $p_T$ distribution at $5.02$~TeV in comparison to $2.76$~TeV, where the IPSAT results improve the data description at $p_T > 100$~GeV in all centrality intervals. Moreover, the predictions with $N_{part}$ and $N_{coll}$ scalings somewhat agree with the experimental measurements at $p_T < 100$~GeV. On the other hand, the estimates based on $N_{coll}$ scaling describes properly the $p_T$ distribution at $2.76$~TeV, whereas the $N_{part}$ scaling is in accordance with measurements for the most central collisions (0--10\%, 10--30\%). The $N_{part}$ scaling is loosing adherence to data in minimum bias case for 2.76~TeV (left panel), whereas at 5.02~TeV the agreement is recovered.

\begin{figure*}[t]
\begin{tabular}{c}
\includegraphics[width=\textwidth]{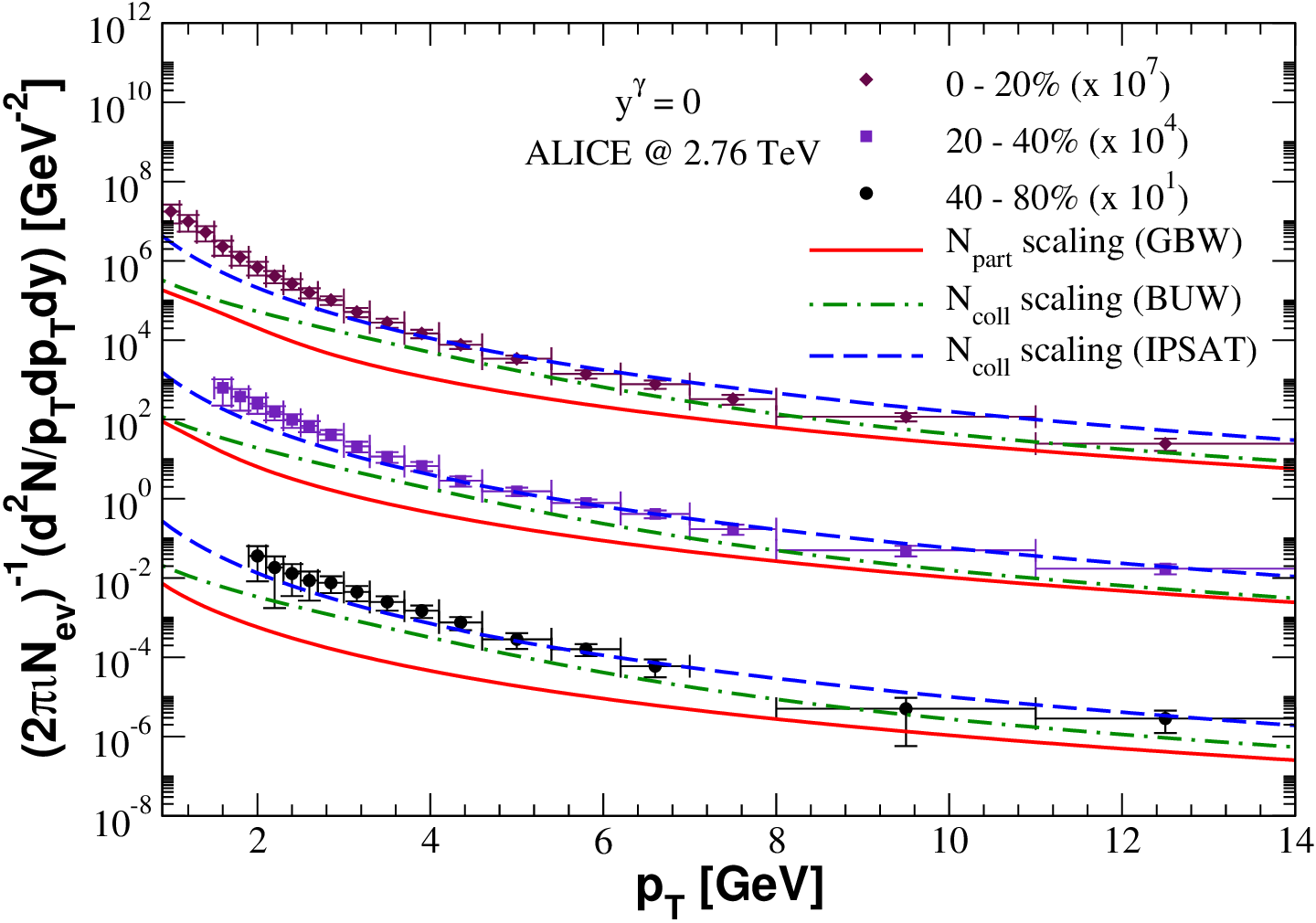}
\end{tabular}
\caption{The $p_T$ distribution for the prompt photon yield in PbPb collisions at $\sqrt{s} = 2.76$~TeV for the central photon   rapidity bin. The results with two distinct geometrical scalings are compared to the experimental measurements reported by the ALICE Collaboration \cite{alice}.}
\label{alice}
\end{figure*}

For the purpose of investigating photon yield production at low $p_T$, we show in Fig.~\ref{alice} our results compared to the measurements provided by the ALICE Collaboration \cite{alice} at $2.76$~TeV in central rapidities. As we can notice in all three centrality cases the $N_{part}(N_{coll})$ results are not consistent with the data in the kinematic range $p_T < 4$~GeV, especially the predictions with $N_{part}$ scaling. The IPSAT and BUW results seem to predict the correct shape of the experimental data as the $p_T$ spectrum increases starting at 4~GeV. One possibility for the underestimation of the experimental measurements at $p_T < 4 $~GeV is that this $p_T$ domain is dominated by contributions from thermalized and hadronic phases at which thermal photons are emitted \cite{kapusta,turbide}. Therefore, the excess of photons verified in this particular $p_T$ region could be associated to thermal photon production, the same situation was found in pQCD calculations \cite{gordon2,vogelsang,paquet} presented in Ref.~\cite{alice}. 

It is timely to discuss the main uncertainties involved in the theoretical predictions. In the $N_{part}$ scaling approach, we are using a particular form for the nuclear saturation scale \cite{salgado} where there is a sizable uncertainty associated to the models for the saturation scaled for nuclei. Moreover, only the GBW dipole-proton amplitude was considered and is known that the specific behavior of the dipole amplitude close to saturation line can be different. We stress that the theoretical predictions are completely parameter free once no fit of parameters has been done. For instance, the parameter $\kappa$ appearing on the saturation scale, Eq.~(\ref{qs2A}), could be fitted from the $p_T$-spectra. As discussed before, the simplified assumption that $\frac{d^3\sigma}{dyd^2p_T}(AA\rightarrow \gamma X)\sim F_2^A(x_1,\mu^2)\otimes \sigma (q+A\rightarrow \gamma X)$ is debatable. Also, the specific model for the nuclear structure function is an additional source of uncertainty. In our case a leading twist shadowing parametrization has been used \cite{f2a}. For the $N_{coll}$ scaling, the number of sources of uncertainties is quite reduced. The effective nuclear dependence is taken into account via Glauber model for hard QCD scattering cross section. The cross section variation using two models for the dipole-proton amplitude (BUW and IPSAT) is not so high and IPSAT does a good job even at low-$p_T$. There are other treatments for the effects of quantum coherence on the $p_T$ distribution of photons radiated by a quark propagating through nuclear matter as done in Refs.~\cite{kop1,Kopeliovich:2001xj,Krelina:2017jfm}, including the next Fock state $q\bar{q}g$ leading to gluon shadowing in color dipole cross section.

In order to single out wether effects of parton saturation effects are relevant to the spectra or not, the nuclear saturation scale is calculated for the energies, rapidity bins and centralities associated to the experimental measurements in Table \ref{Tab:1}. We select some representative samples of the three experimental kinematic ranges. One sees that $Q_{sA}^2\lesssim 1$ GeV$^2$ for any centrality interval, with the higher values corresponding to more central collisions. The smallness of the nuclear saturation scale is due to the lower value of the center-of-mass energy compared to $pp$ collisions and the large $p_T$ measured. In any case, $p_T^2\gg Q_{sA}^2$ and interactions are dominated by the  color transparency regime of the dipole-nucleus cross section. Of course, the situation changes when small $p_T$ is considered as shown in Refs.~\cite{kbml,pras,kp}, where the saturation effects are enhanced.

 \begin{table*}[!bh]
\centering
\caption{The typical values for the nuclear saturation scale, $Q^{2}_{sA}(x_2)$, for the average kinematic range of measured prompt photon $p_T$ spectra considering some representative centrality ranges.}
\begin{tabular}{l|c|c|c|c}
\hline
  $\sqrt{s_{AA}}$ (TeV) & $\langle y \rangle$ and $\langle p_T \rangle$ bins & Centrality & $N_{part}$ & $Q^{2}_{sA}$ (GeV$^2$) \\
\hline
 2.76 (ATLAS) & 0 (1.95) and 150 (80) GeV   & 0 -- 10\%  & 356.2 & 0.550 (1.05)  \\
 2.76 (ATLAS) & 0 (1.95) and 150 (80) GeV  & 40 -- 80\% & 45.9  & 0.041 (0.078)  \\
 \hline 
 5.02 (CMS) & 0  and 112.5 GeV   & 0 -- 10\%  & 356.2 & 0.690   \\
 5.02 (CMS) & 0 and 112.5  GeV  & 50 -- 100\% & 21.9  & 0.020  \\
  \hline
 2.76  (ALICE) & 0  and 7.5 GeV   & 0 -- 20\%  & 308.0 & 0.970   \\
2.76  (ALICE) & 0 and 7.5 GeV  & 40 -- 80\% & 45.9  &  0.087 \\
 \hline
\end{tabular}
\label{Tab:1}
\end{table*}
 
\begin{figure*}[t]
\begin{tabular}{c}
\includegraphics[width=\textwidth]{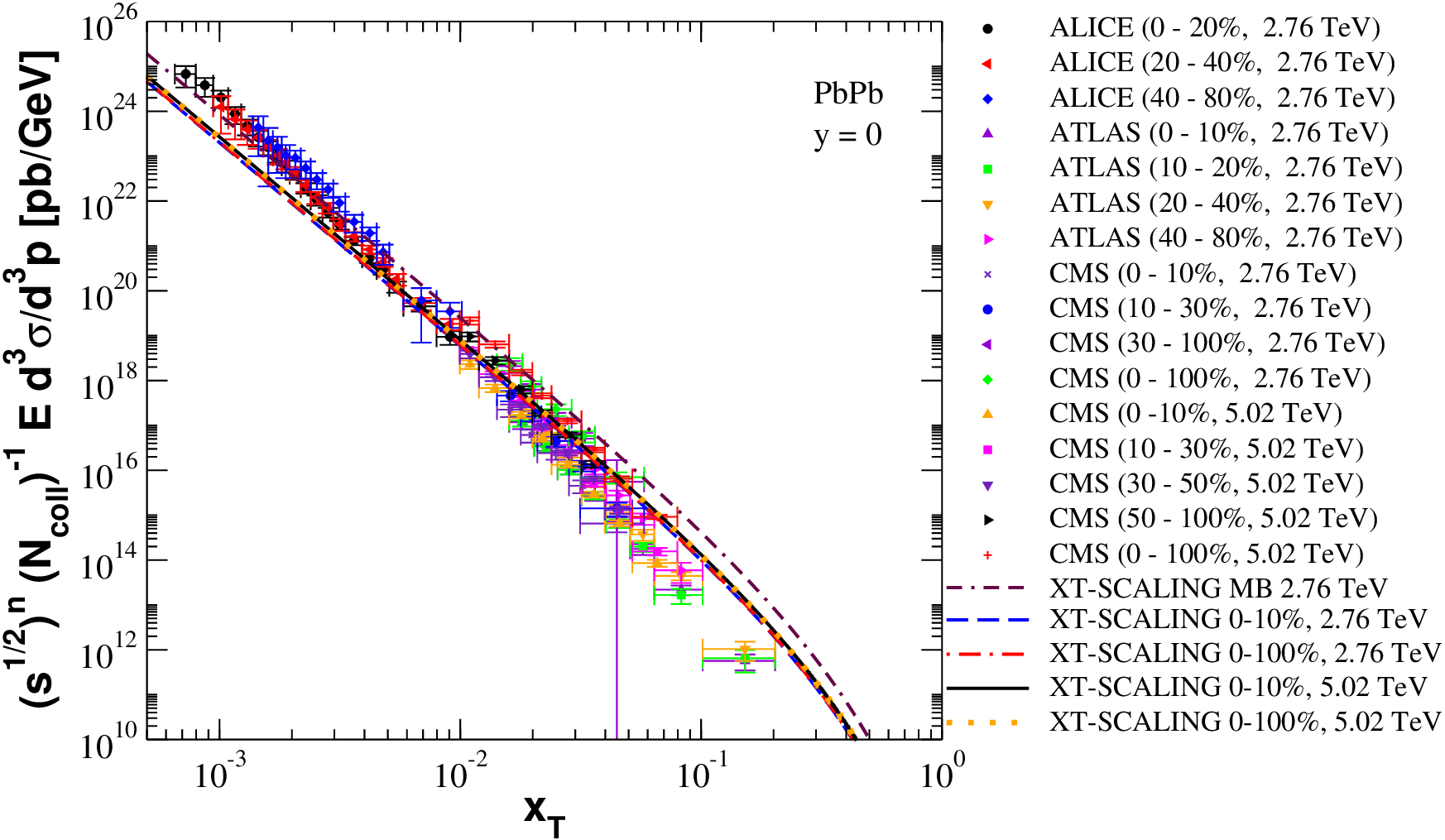}
\end{tabular}
\caption{The $x_T$-scaling observed in prompt photon production in PbPb collisions at central rapidity considering $\sqrt{s} = 2.76$ and 5.02~TeV. The results from analytic expressions, Eqs.~(\ref{xtAA}) and (\ref{xtAANPART}), are shown take into account the more central (0 -- 10\%) and peripheral (0 -- 100\%) centrality classes.}
\label{xtscaling}
\end{figure*}

Although parton saturation does not suppress the spectra in the present kinematic range, the geometric scaling property associated to saturation formalism has consequences on scaling property of the $AA$ cross section. A scaling behavior appears in the saturation formalism considering our assumption that the $AA$ cross section is the convolution of the nuclear structure function with the partonic quark-nucleus cross section. Based on the analysis done in \cite{salgado} in the minimum bias case, one may assume $F_2^A(x_1,\mu^2)\sim \frac{S_A}{S_p}Q_{sA}^2$ at sufficiently small $x_1$. On the other hand, the partonic cross section contains the dipole-nucleus cross section, which presents geometric scaling property as well. Namely, $\sigma_{dip}^{nuc}\sim \frac{S_A}{S_p}N(Q_{sp}^2\rightarrow Q_{sA}^2)$. Let us take the nuclear structure function having a scaling form:
\begin{eqnarray}\nonumber
F_2^A(x,Q^2) &\approx& \frac{\sigma_0 Q^2}{4\pi^2\alpha_{em}}\frac{S_A}{S_p}\left(\frac{Q_{sA}^2(x)}{Q^2}\right)\simeq \frac{S_A}{S_p}\kappa(A)F_2^p(x,Q^2),\\ \\
\kappa(A) &=& \left(\frac{AS_p}{S_A}  \right)^{\Delta}
\end{eqnarray}
where the nuclear saturation scale is set to $Q_{sA}^2(x)=Q_{sp}^2(x)\kappa(A)$ as in Eq.~(\ref{qs2A}) and with $\Delta =1/\delta \simeq 1.27$. We have shown in Refs.~\cite{gsds,gsds1} that a $x_T$-scaling ($x_T = 2\,p_T/\sqrt{s}$) expression can be obtained for the $pp$ case,
\begin{eqnarray}
E\frac{d^{3}\sigma^{pp\to \gamma X}}{d^3p}(x_T) & \approx & \frac{N_0}{\big(\sqrt{s}\big)^{4}}\,\bigg(\frac{x_T}{2}\bigg)^{-n}\,G(x_1),
\label{xtpp}
\end{eqnarray}
with $n\simeq 4.5$ and $G(x_1)$ being a well behaved function of $x_1=(x_T/2)e^y$. Furthermore, the overall normalization is given by $N_0 = \bar{\sigma}_{pp}\,(x_0)^{2\,\lambda}$. For $AA$ collisions based on geometric scaling~\cite{salgado}, the invariant cross section can be expressed as
\begin{eqnarray}\nonumber
E\frac{d^{3}\sigma}{d^3p}^{AA\to \gamma X} & \approx & \frac{N_0}{\big(\sqrt{s}\big)^{4}}\,\bigg(\frac{S_A}{S_p}\bigg)^2\,
\bigg(\frac{A\,S_p}{S_A}\bigg)^{2\Delta}\,\bigg(\frac{x_T}{2}\bigg)^{-n}\,G(x_1), \\
& \approx & A^2\bigg(\frac{A\,S_p}{S_A}\bigg)^{\varepsilon}E \frac{d^{3}\sigma}{d^3p}^{pp\to \gamma X}(x_T),
\label{xtAA}
\end{eqnarray}
where $\varepsilon=2(1-\delta)/\delta \simeq 0.54$. This theoretical prediction can be compared to the experimental measurements. In Fig.~\ref{xtscaling} a compilation of prompt photon $PbPb$ data at midrapidity at the LHC energies is presented and the dash-dotted line represents the result based in Eq.~(\ref{xtAA}). The phenomenological parametrization above is related to minimum bias configuration. The case for the energy of 2.76~TeV is presented (label {\tt{XT-SCALING MB 2.76 TeV}}), which is quite consistent with data for $x_T\leq 10^{-2}$ and overestimate them for larger $x_T$. However, we can write down an expression in terms of $N_{part}$ using the calculation for the nuclear saturation scale in Eq.~(\ref{qs2A}),
\begin{eqnarray}\nonumber
E\frac{d^{3}\sigma}{d^3p}^{AA\to \gamma X} &\approx& 
A^2\bigg(\frac{A\,S_p}{S_A}\bigg)^{\varepsilon} \\\nonumber
&\times& \left(\frac{\kappa(b) N_{part}}{A}  \right)^{\frac{\varepsilon}{2}+1} E\frac{d^{3}\sigma}{d^3p}^{pp\to \gamma X}(x_T).\\
\label{xtAANPART}
\end{eqnarray}
The parametrization above is shown in Fig.~\ref{xtscaling} for some particular centrality intervals assuming $\sqrt{s} = 2.76$~TeV (labeled dashed and dot-dashed curves) and $\sqrt{s} = 5.02$~TeV (solid and dotted curves). One sees that results are almost independent of centrality and energy, in reasonable agreement with the experimental scaling plot. The prediction describes the experimental measurements in the range $10^{-2}\lesssim x_T\lesssim 10^{-1}$. It should be stressed that analytical results in Eqs.~(\ref{xtAA}) and (\ref{xtAANPART}) are qualitative due to the many approximations an assumptions made. In any case, the centrality dependence can be absorbed in the nuclear saturation scale leading to $x_T$ scaling shown by data.

\section{Summary} 
\label{conc}

In this work we performed an analysis on photon yield production in PbPb collisions at the LHC energies within the color dipole formalism. We demonstrate that the prompt photon yield exhibits geometric scaling rules which can be properly accounted by considering the color dipole framework. The $N_{part}$ and $N_{coll}$ results are able to reasonably describe the data at large $p_T$ in comparison to the data reported by the ATLAS, ALICE and, CMS Collaborations, in particular better results are provided by $N_{coll}$ scaling predictions. Moreover, our results are not in agreement to the photon yields measured at low $p_T$ spectra, where contributions from thermal photons are expected and are not considered in our calculations that accounts hard direct photon production.

The quantities $N_{part}$ and $N_{coll}$ are associated to the total number of nucleons that effectively participate of a collision and the total number of nucleon collisions, respectively. The first is assumed to determine the collision centrality and the latter is related to the production yield, being properly to scaling the PbPb yield, which we are considering within the color dipole model providing consistent results at high $p_T$, region that apparently is dominated for hard direct photon.

Additionally, we have demonstrated that, within the color dipole picture, we can derive phenomenological parametrizations for the prompt photon $x_T$-scaling associated to minimum bias and $N_{part}$ configuration in $AA$ collisions. Such parametrizations are consistent with the data at small $x_T$, i.e, low $p_T$ distribution, which is significant given the simple form of the related analytical expressions that can be convenient in future investigations of experimental measurements of prompt photons in heavy-ion collisions.

\section*{Acknowledgements}

This work was partially financed by the Brazilian funding agencies CAPES, CNPq, and FAPERGS. This study was financed in part by the Coordena\c{c}\~ao de Aperfei\c{c}oamento de Pessoal de N\'{\i}vel Superior - Brasil (CAPES) - Finance Code 001. GGS
acknowledges funding from the Brazilian agency Conselho Nacional
de Desenvolvimento Científico e Tecnológico (CNPq) with grant
CNPq/313342/2017-2.

\end{document}